\documentclass[11pt]{article}

\usepackage[final]{acl}
\usepackage{subcaption, algorithm2e, url, float, etoolbox, adjustbox, pgf,booktabs, verbatim}
\usepackage{xcolor}
\usepackage{amsmath}
\usepackage{amssymb}
\usepackage{tcolorbox}
\usepackage{multirow}
\usepackage[table,xcdraw]{xcolor}
\usepackage{times}
\usepackage{latexsym}
\usepackage{hyperref}
\usepackage{xcolor}

\usepackage[T1]{fontenc}

\usepackage[utf8]{inputenc}

\usepackage{microtype}

\usepackage{inconsolata}

\usepackage{graphicx}

\title{\texttt{Indic-CodecFake meets SATYAM:} Towards Detecting Neural Audio Codec Synthesized Speech Deepfakes in Indic Languages}

\author{
\textbf{Girish\textsuperscript{1}\thanks{Equal contribution as first author.}} \quad
\textbf{Mohd Mujtaba Akhtar\textsuperscript{2}\footnotemark[1]} \quad
\textbf{Orchid Chetia Phukan\textsuperscript{3}\footnotemark[1]\thanks{Core Ideation}} \quad
\textbf{Arun Balaji Buduru\textsuperscript{3}}\\
\textsuperscript{1}UPES, India \quad
\textsuperscript{2}Veer Bahadur Singh Purvanchal University, India \quad
\textsuperscript{3}IIIT-Delhi, India\\
\texttt{\textbf{Correspondence:}\textcolor{blue}{ orchidp@iiitd.ac.in}}
}

\begin{document}
\maketitle
\begin{abstract}
The rapid advancement of Audio Large Language Models (ALMs), driven by Neural Audio Codecs (NACs), has led to the emergence of highly realistic speech deepfakes, commonly referred to as CodecFakes (CFs). Consequently, CF detection has attracted increasing attention from the research community. However, existing studies predominantly focus on English or Chinese, leaving the vulnerability of Indic languages largely unexplored. To bridge this gap, we introduce Indic-CodecFake (ICF) dataset, the first large-scale benchmark comprising real and NAC-synthesized speech across multiple Indic languages, diverse speaker profiles, and multiple NAC types. We use IndicSUPERB as the real speech corpus for generation of ICF dataset. Our experiments demonstrate that state-of-the-art (SOTA) CF detectors trained on English-centric datasets fail to generalize to ICF, underscoring the challenges posed by phonetic diversity and prosodic variability in Indic speech. Further, we present systematic evaluation of SOTA ALMs in a zero-shot setting on ICF dataset. We evaluate these ALMs as they have shown effectiveness for different speech tasks. However, our findings reveal that current ALMs exhibit consistently poor performance. To address this, we propose \textbf{\texttt{SATYAM}}, a novel hyperbolic ALM tailored for CF detection in Indic languages. \textbf{\texttt{SATYAM}} integrates semantic representations from Whisper and prosodic representations from TRILLsson using through Bhattacharya distance in hyperbolic space, and subsequently performs the same alignment procedure between the fused speech representation and a input conditioning prompt. This dual-stage fusion framework enables \textbf{\texttt{SATYAM}} to effectively model hierarchical relationships both within speech (semantic–prosodic) and across modalities (speech–text). Extensive evaluations show that \textbf{\texttt{SATYAM}} consistently outperforms competitive end-to-end and ALM-based baselines on the ICF benchmark.
\end{abstract}

\section{Introduction}

Speech deepfakes have rapidly transitioned from proof-of-concept demonstrations to tools deployed in large-scale, real-world attacks. Recent years have witnessed a surge in financial frauds \footnote{\href{https://red-goat.com/voice-cloning-heist/}{Red Goat}}  
\footnote{\href{https://cyfor.co.uk/cybercriminals-clone-voice-of-company-director-in-35-million-bank-heist/}{Voice Cloning Heist}}. With only a few seconds of recorded speech, attackers can now generate long, natural-sounding utterances that preserve a target speaker’s accent, prosody, and speaking style~\cite{khanjani2023audiodeepfakesurvey}. Beyond financial fraud, such synthesized speech has also been exploited for spreading disinformation and manipulating public opinion~\cite{luong2025llamapartialspoof}. These fake voices are predominantly generated by text-to-speech (TTS) models and voice conversion (VC) techniques. Modern TTS architectures such as WaveNet and VITS can synthesize expressive speech directly from text~\cite{vandenoord16_ssw,kim2021vits,tan2021ttssurvey}, while VC models such as AutoVC and StarGAN-VC2 enable many-to-many voice style transfer without requiring parallel data~\cite{qian2019autovc,kaneko2019starganvc2}. \par
In response to these threats, the speech community has devoted substantial efforts towards building effective systems for detecting such fakes. Benchmark series such as ASVspoof~\cite{wu2015asvspoof, asvspoof2019,asvspoof2021, wang2024asvspoof} have driven progress on spotting synthetic and replayed speech, and a wide range of countermeasures have been explored. Early systems rely on handcrafted spectral features with classical machine learning techniques such as GMM, Random Forest and so on \cite{patel2015combining, patel2016cochlear, yu2017spoofing, ji2017ensemble}. Subsequent studies have leveraged deep learning algorithms for improved speech deepfake detection performance \cite{lei2020siamese, dinkel2017end, jung2022aasist, huang-etal-2025-speechfake}. Building on this research landscape, the start of this decade marked a fundamental shift in speech deepfake detection with the widespread adoption of large-scale pre-trained models (PTMs) trained on diverse and extensive speech corpora.  As a result, recent studies have evaluated a variety of state-of-the-art (SOTA) PTMs—including WavLM, Wav2Vec2, Whisper, and MMS—for speech deepfake detection~\cite{kawa23b_interspeech, phukan2024heterogeneity, muller2024mlaad, el2025comprehensive}. More recently, audio large language models (ALMs), such as Qwen2-Audio, have also been explored for speech deepfake detection, motivated by their strong performance across a range of related speech processing tasks~\cite{10.1145/3746027.3755851}. \par 

\begin{figure}[!bt]
    \centering
    \includegraphics[width=1\linewidth]{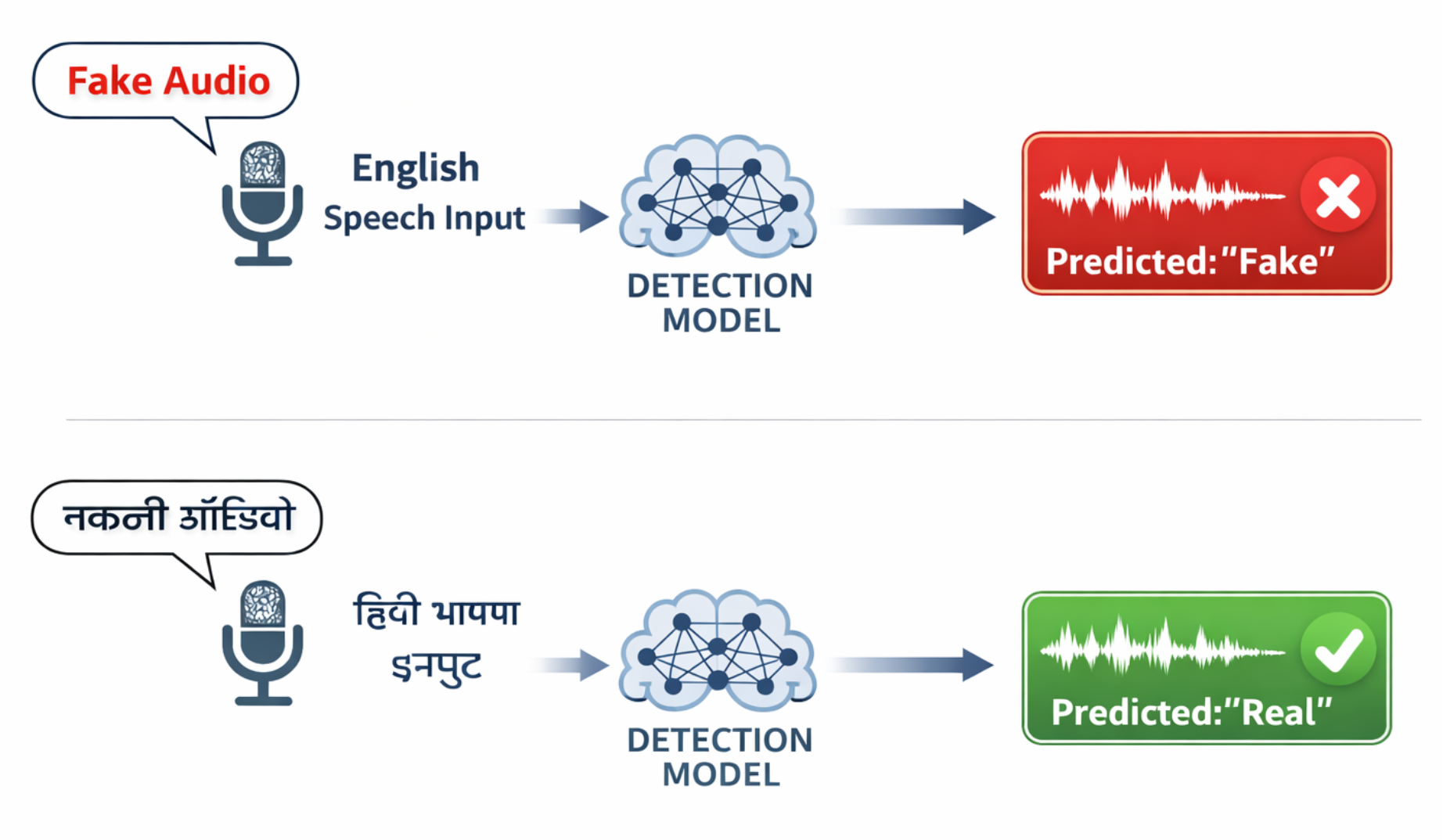}
    \caption{Existing CodecFake detectors perform well on English speech but frequently misclassify Hindi fake speech as real}
    \label{fig:hindieng}
\end{figure}

However, the majority of these studies primarily focus on speech deepfakes generated by TTS or VC systems. In recent years, driven by advances in ALMs, a new category of speech deepfakes—referred to as CodecFakes (CFs)—has emerged and attracted growing attention from the research community. These ALMs incorporate neural audio codecs (NACs) within their architectures for both speech encoding and synthesis. For instance, AudioLM relies on the SoundStream as the backbone NAC ~\cite{borsos2023audiolm}. In response to this emerging threat, \cite{wu24p_interspeech} and \cite{lu24f_interspeech} initiated the first explorations of CF detection by releasing dedicated CF datasets. Their studies demonstrate that models trained on traditional speech deepfake datasets fail to generalize effectively to CF benchmarks, largely due to shifts in the underlying distributional characteristics of CFs compared to TTS- or VC-generated synthetic speech. As such various succeeding studies have been conducted in this direction for CF detection \cite{xie2025codecfake, chen2025codecfake+}. Despite this progress, existing CF detection datasets are predominantly English-focused or atmost including Chinese. However, CF datasets for other language groups remains largely unaddressed. \par
In this work, we extend CF detection to Indic languages, one of the world’s most linguistically diverse settings. India, now the most populous country globally encompasses speakers from major Indo-European and Dravidian families as well as Austro-Asiatic and Tibeto–Burman languages. This diversity has fueled rapid growth in Indic speech technologies, including large-scale benchmarks such as IndicSUPERB~\cite{javed2023indicsuperb}. At the same time, surveys report widespread exposure to AI-driven voice scams~\footnote{
\href{https://www.mcafee.com/content/dam/consumer/en-us/resources/cybersecurity/artificial-intelligence/rp-beware-the-artificial-impostor-report.pdf}{McAfee, \emph{Beware the Artificial Impostor} (report)}.}
. Despite this elevated risk, existing CF datasets remain largely English-centric. To address this gap, we introduce Indic-CodecFake (ICF) dataset, the first large-scale benchmark comprising real and NAC-synthesized speech across multiple Indic languages, diverse speaker profiles, and multiple NAC types. We show that models trained on previous CF detection benchmarks fails on ICF dataset (Figure \ref{fig:hindieng}). We further conduct a systematic zero-shot evaluation of SOTA ALMs on ICF dataset. We evaluate ALMs as previous research has explored them for speech deepfake detection but they haven't evaluated them for CF detection \cite{10.1145/3746027.3755851}. We observe consistently poor performance on ICF dataset and thus showing the need for Indic-centric CF detection pipelines. To overcome these limitations, we propose \textbf{\texttt{SATYAM}}, a hyperbolic ALM tailored for Indic CF detection. \textbf{\texttt{SATYAM}} leverages Bhattacharya distance in hyperbolic space to align semantic representations from Whisper with prosodic representations from TRILLsson, followed by a second stage of the same alignment with a input conditioning prompt. This hierarchical modeling enables effective integration of speech–speech and speech–text relationships. Extensive experiments demonstrate that \textbf{\texttt{SATYAM}} consistently outperforms competitive ALM-based and end-to-end baselines on ICF, while also achieving strong performance on existing CF detection benchmarks. \par

\noindent \textbf{To summarize, the major contributions are as follows}

\begin{itemize}
    \item We introduce ICF dataset, the first large-scale CF dataset in Indic-languages. 
    \item We evaluate previous SOTA CF detectors trained on previous CF benchmarks on ICF dataset and show its poor generalization. We also present a evaluation of SOTA ALMs on ICF dataset and observe poor performance. This highlights the need for Indic-centric CF detection modeling pipelines. 
    \item We propose, \textbf{\texttt{SATYAM}}, a novel hyperbolic ALM for CF detection primarily in Indic Languages. To the best of our knowledge, we are the first study to explore extension of ALMs to hyperbolic space.
\end{itemize}

\noindent \textit{We release the dataset, data generation pipeline and code here\footnote{\url{https://helixometry.github.io/IndicFake/}}.}

\section{Related Works}


\noindent In this section, we briefly review prior work on CF detection. Early investigations into CF detection and its associated vulnerabilities were initiated by Wu et al.~\cite{wu24p_interspeech} and Lu et al.~\cite{lu2024codecfake}. These studies demonstrated that SOTA speech deepfake detectors trained on traditional deepfake datasets—primarily synthesized using TTS or VC systems—fail to generalize effectively to CF scenarios. Wu et al.~\cite{wu24p_interspeech} constructed a CF dataset using the English VCTK corpus and a diverse set of NACs, and evaluated AASIST-based architectures for CF detection. In parallel, Lu et al.~\cite{lu2024codecfake} and Xie et al.~\cite{xie2025codecfake} developed CF datasets spanning English and Chinese by leveraging VCTK and AISHELL-3 as real speech corpora, respectively, and similarly adopted AASIST-based modeling approaches. Building on these efforts, Du et al.~\cite{xie2025codecfake} further expanded the CF detection landscape by incorporating a larger variety of NAC families while continuing to rely on AASIST-based architectures. In contrast, Xie et al.~\cite{xie2025codecfake} proposed a novel CF detection strategy based on sharpness-aware minimization to improve robustness. Despite this progress, existing CF datasets are almost exclusively limited to high-resource languages, primarily English and Chinese, underscoring the need for multilingual CF benchmarks. In particular, no prior work has systematically focused on CF detection in Indic languages. While Cui et al.~\cite{cui2025whiadd} explored multilingual CF detection by building upon the Common Voice corpus and included Tamil as an Indic language, their dataset is not publicly released and considers only a single Indic language. In addition, several Indic-language–focused datasets have been proposed for general speech deepfake detection mostly TTS, VC~\cite{sharma2025indicsynth,ranjan2025indicfake}; however, none specifically target CF detection. Prior studies have shown that Bhattacharyya distance is effective for speech representation alignment, while geometry-aware modeling has also shown promise for improving generalization in ADD \cite{phukan25d_interspeech,sheth-etal-2025-curved}. In this work, we address this gap by introducing ICF dataset, the first large-scale and comprehensive CF dataset focused on Indic languages. Furthermore, we propose \textbf{\texttt{SATYAM}}, a novel hyperbolic ALM tailored for CF detection in Indic languages.

\section{Indic-CodecFake Dataset}
\label{sec:indic-codecfake}

In this section, we first describe the real-speech source datasets for Indic languages, followed by the SOTA NACs considered in our study. We then detail the data generation pipeline used to construct the ICF dataset. 

\noindent \textbf{Indic Speech Source}: We use the IndicSUPERB\footnote{\url{https://github.com/AI4Bharat/IndicSUPERB?tab=readme-ov-file}} \cite{javed2023indicsuperb} dataset as the real speech corpus. It consists of 12 Indian-languages. IndicSUPERB is also used by IndicSynth \cite{sharma2025indicsynth} as a base real speech corpus. IndicSynth is a SOTA large-scale speech deepfake benchmark on Indic-languages generated through TTS or VC methods. Additional details about IndicSUPERB is given in Appendix \ref{sec:appendix} IndicSUPERB.

\noindent \textbf{Neural Audio Codecs}: We follow prior work by \citet{lu24f_interspeech} and \citet{wu24p_interspeech} and adopt SOTA publicly released NACs that are widely available and easy to reproduce. \textbf{DAC} \cite{kumar2024high}: We use 16\,kHz, 24\,kHz, and 44\,kHz variants. \textbf{Encodec} \cite{defossez2022high}: We use 24\,kHz and 48\,kHz models. \textbf{SoundStream} \cite{zeghidour2021soundstream}:  We use the 16\,kHz configuration. \textbf{SpeechTokenizer} \cite{zhang2024speechtokenizer}: We use the default 16\,kHz setup. \textbf{FunCodec} \cite{du2024funcodec}: We use the official 16\,kHz version. \textbf{AudioDec} \cite{wu2023audiodec}: We use 28\,kHz and 48\,kHz variants. \textbf{SNAC}\ \cite{siuzdak2024snac}: We use 24\,kHz, 32\,kHz, and 44\,kHz models. \textbf{MIMI} \cite{defossez2024moshi}: It operates in 24 kHz. To support reproducibility, we provide a repository of the NAC resources used in data generation.\footnote{\url{https://github.com/CodeVault-girish/Neural-Codecs.git}}

\begin{figure*}[hbt!]
\centering
    \includegraphics[width=0.971\linewidth]{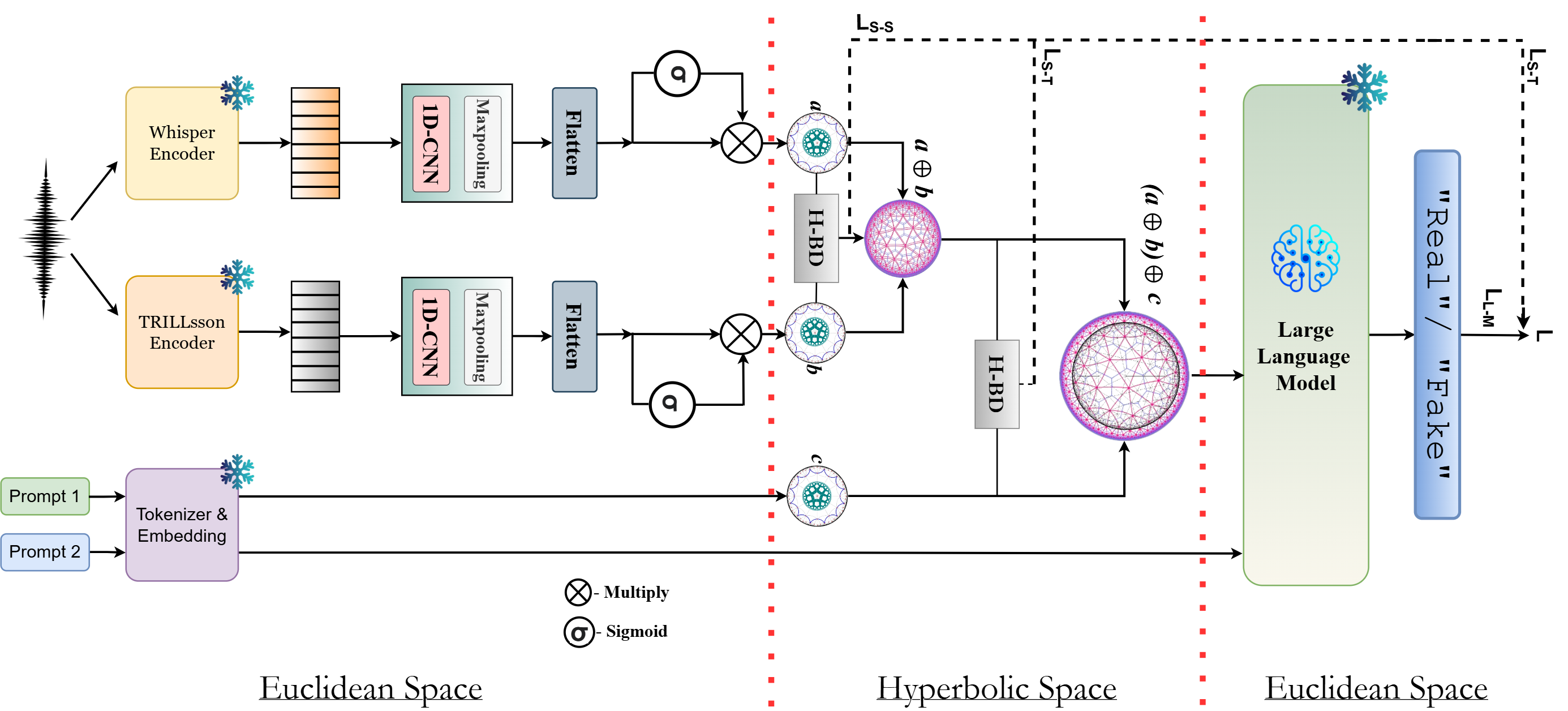}
    \caption{Proposed Framework: \textbf{\texttt{SATYAM}}; H-BD stands for Bhattacharya distance in Hyperbolic space}
    \label{archi}
\end{figure*}

\noindent \textbf{Generation Pipeline of ICF}: To construct ICF, we follow a controlled NAC-based resynthesis pipeline inspired by prior work on CF generation~\cite{wu24p_interspeech}. Specifically, we resynthesize real speech from IndicSUPERB using a diverse set of NACs. All source utterances are drawn directly from the official IndicSUPERB train/valid/test splits and treated as real references. Given a real speech waveform $x$, a NAC encoder $\mathcal{E}$ maps it to a discrete latent representation $z=\mathcal{E}(x)$, which is then reconstructed by a decoder $\mathcal{D}$ to obtain $\tilde{x}=\mathcal{D}(z)$. The reconstructed signal $\tilde{x}$ constitutes the corresponding CF sample. This process of encoding-decoding preserves linguistic content and speaker characteristics while introducing NAC-specific artifacts representative to each NAC. We apply this procedure to every utterance and each selected NAC, resulting in parallel corpora where each real speech sample has a one-to-one CF counterpart for every NAC configuration. We retain the original split assignment given in IndicSUPERB during the CF generation process.  We define two evaluation settings: (i) Seen: Training and evaluation involve CFs synthesized using the same set of NACs (e.g., SNAC variants, DAC variants, Encodec, SoundStream, and SpeechTokenizer). As IndicSUPERB, contains two sets of test set, we use the test-known for this evaluation setting. (ii) Unseen: The test set contains CFs generated by NACs not observed during training (e.g., FunCodec variants, AudioDec variants, and MIMI), enabling evaluation of cross-codec generalization. We use the test-unkown test split for this evaluation setting. This pipeline is applied uniformly across all Indic languages in IndicSUPERB.

\section{Methodology}
\label{satyam}

In this section, we describe the methodology underlying the proposed hyperbolic ALM, \textbf{\texttt{SATYAM}}. An overview of the model architecture is illustrated in Figure~\ref{archi}.  We propose \textbf{\texttt{SATYAM}}, a supervised hyperbolic ALM for CF detection that formulates detection as a conditional generation task, following prior ALM-based approaches for speech deepfake detection ~\cite{10.1145/3746027.3755851}. They formulate speech deepfake detection as a audio question answering task and has shown its effectiveness. We begin by describing the audio encoders and LLM decoder employed in our approach. We then present the complete workflow of \textbf{\texttt{SATYAM}}, including hyperbolic fusion of semantic and paralinguistic speech representations and their alignment with prompt inputs. Our design is motivated by recent findings showing that audio encoders constitute the primary performance bottleneck in existing ALM pipelines, and that employing stronger, task-relevant speech encoders leads to substantial gains in downstream tasks~\cite{li2025dfallm}. While prior work has primarily focused on general speech deepfake detection, the potential of such encoder-centric ALM designs for CF detection remains largely unexplored. Furthermore, we adopt hyperbolic geometry for both speech--speech fusion and speech--prompt alignment based on the observation that semantic and paralinguistic cues might exhibit inherent hierarchical structure \cite{mary2015multiple, chen2023speechformer++}. Prior work by Phukan et al.~\cite{phukan25b_interspeech} has shown evidence of hierarchical organization within speech representations, while hierarchical relationships across modalities such as speech, vision and text are a well-established principle in multimodal learning~\cite{desai2023hyperbolic, hong2023hyperbolic}. Hyperbolic space, with its ability to naturally embed hierarchical structures, therefore provides a principled geometric framework for modeling these relationships in \textbf{\texttt{SATYAM}}. 
\par

\noindent
\textbf{Audio Encoders}: We use Whisper \cite{radford2023robust} and TRILLsson ~\cite{shor2022trillsson} as audio encoders as they have proven effective by previous research on synthetic speech detection \cite{hetero2024, das25_interspeech}. More information about the audio encoders are given in Appendix \ref{sec:appendix} Additional Information about Audio Encoders.

\noindent \textbf{LLM decoder}: We employ Qwen2-7B\footnote{\url{https://huggingface.co/Qwen/Qwen-7B}} as the pre-trained decoder-only LLM backbone in our framework~\cite{team2024qwen2}. It is selected as it has demonstrated effectiveness across a range of speech and audio language modeling tasks, including speech deepfake detection~\cite{li2025dfallm} and speech emotion recognition~\cite{su2025reasoning}. Furthermore, Qwen2-Audio, a SOTA ALM, is explicitly built upon the Qwen2 decoder architecture, further validating the suitability of Qwen2 as a strong LLM backbone for speech and audio tasks~\cite{chu2024qwen2}.

\noindent
\textbf{Workflow:}  Given an input speech utterance $x$, the objective of \textbf{\texttt{SATYAM}} is to determine whether the utterance is real or fake by generating a short natural-language decision. From $x$, we extract two complementary speech representations: a semantic representation using Whisper, $e_w \in \mathbb{R}^{d_w}$, and a paralinguistic representation using TRILLsson, $e_t \in \mathbb{R}^{d_t}$. Both representations are passed through a lightweight CNN block consisting of a 1D convolutional layer (filter size $3$) followed by max pooling, as adopted in prior work~\cite{hetero2024}. The resulting representations are then projected into a shared Euclidean space of dimension $d$: $ e_w = W_w e_w $ and $e_t = W_t e_t $ where $W_w$ and $W_t$ are learnable projection matrices. We further introduce a sigmoid gating module that filters the input representations signal and forwards only salient information to the subsequent stage. We next map these representations into a $d$-dimensional hyperbolic space $\mathbb{H}_c^d$ with curvature $-c$, where $c>0$. The Euclidean representations are mapped to the hyperbolic manifold using the exponential map at the origin:
\[
\exp_0^c(u)
=
\tanh(\sqrt{c}\|u\|)
\frac{u}{\sqrt{c}\|u\|}
\]
This yields hyperbolic speech representations $ h_w = \exp_0^c(\tilde e_w) $ and $ h_t = \exp_0^c(\tilde e_t)$ with $h_w, h_t \in \mathbb{H}_c^d$. We then proceed on to fusing the semantic and paralinguistic hyperbolic speech representations. To align semantic and paralinguistic speech representations, we minimize the Bhattacharyya distance (BD) (BD has shown effectiveness in aligning speech representations \cite{phukan25d_interspeech}), however, it was for euclidean space and we extend it to hyperbolic space in this work) between the corresponding hyperbolic distributions. Lower BD represents greater alignment and so we aim to optimize it to minima. For two feature distributions $P$ and $Q$ on $\mathbb{H}_c^d$, the BD is defined as
\[
D_B(P,Q)
=
-\log
\int_{\mathbb{H}_c^d}
\sqrt{p(h)\,q(h)}\, d\mu_c(h)
\]
The speech--speech alignment loss is given by $\mathcal{L}_{S\text{-}S}
=
D_B\!\left(h_w, h_t\right) $. After aligning the two speech views, we obtain a single fused speech representation using mobius addition, which preserves the geometry of hyperbolic space. For two points $x,y \in \mathbb{H}_c^d$, mobius addition is defined as
\[
\small
x \oplus_c y
=
\frac{
(1 + 2c \langle x, y \rangle + c \|y\|^2)x
+
(1 - c \|x\|^2)y
}{
1 + 2c \langle x, y \rangle + c^2 \|x\|^2 \|y\|^2
}
\]
Using this operation, we compute the fused speech embedding $h_f = h_w \oplus_c h_t $. We feed a conditioning prompt (\textit{``Analyze the speech for unnatural artifacts''}) to emphasize on unnatural artifacts relevant for synthetic speech. Using Qwen-2, we extract hidden states from an intermediate transformer layer. A single prompt representation $e_A$ is obtained via mean pooling over tokens and projected into the shared space.
To inject task-level reasoning, we align the fused speech distribution with the prompt representation using the same BD after we transform the prompt representation to hyperbolic space using exponential map at the origin. $ \mathcal{L}_{S\text{-}T} = D_B\!\left(h_f, h_A\right) $. Following this, we aggregate the aligned hyperbolic representations using mobius addition: $ h_{\mathrm{final}} = h_f \oplus_c h_A $. The aggregated representation is mapped back to Euclidean space using the logarithmic map, $ u_{\mathrm{final}} = \log_0^c(h_{\mathrm{final}}) $ and 
$ g = W_g u_{\mathrm{final}} $ where $g$ is injected as prefix conditioning tokens into a Qwen-2 LLM decoder. Logarithmic map is given by:
\[
\log_0^c(h)=\frac{1}{\sqrt{c}}\tanh^{-1}(\sqrt{c}\|h\|)\frac{h}{\|h\|}
\]

\begin{table*}[hbt!]
\centering
\scriptsize
\setlength{\tabcolsep}{4pt}
\renewcommand{\arraystretch}{1.15}
\caption{Zero-shot evaluation of Qwen2-audio family on ICF and Codecfake \cite{wu24p_interspeech}; ACC stands for Accuracy; All the scores are in \%}
\label{tab:prompt_wise_template}

\begin{tabular}{l|cc|cc|cc|cc|cc|cc}
\toprule
\multicolumn{1}{c}{} & \multicolumn{10}{c}{\textbf{ICF}} \\
\midrule
\textbf{Method}
& \multicolumn{2}{c|}{\textbf{Prompt1}}
& \multicolumn{2}{c|}{\textbf{Prompt2}}
& \multicolumn{2}{c|}{\textbf{Prompt3}}
& \multicolumn{2}{c|}{\textbf{Prompt4}}
& \multicolumn{2}{c|}{\textbf{Prompt5}}
& \multicolumn{2}{c}{\textbf{Prompt 6}} \\
& \textbf{ACC \(\uparrow\)} & \textbf{EER \(\downarrow\)}
& \textbf{ACC \(\uparrow\)} & \textbf{EER \(\downarrow\)}
& \textbf{ACC \(\uparrow\)} & \textbf{EER \(\downarrow\)}
& \textbf{ACC \(\uparrow\)} & \textbf{EER \(\downarrow\)}
& \textbf{ACC \(\uparrow\)} & \textbf{EER \(\downarrow\)}
& \textbf{ACC \(\uparrow\)} & \textbf{EER \(\downarrow\)} \\
\midrule
Qwen2 audio-chat & 11.29 & 89.43 & 11.70 & 89.26 & 12.05 & 88.95 & 11.04 & 89.60 & 10.97 & 90.02 & 11.41 & 89.45 \\
Qwen2 audio-base & 11.20 & 88.99 & 13.35 & 88.66 & \textbf{13.41} & \textbf{88.57} & 12.68 & 89.13 & 12.93 & 89.74 & 12.71 & 89.02 \\
\midrule

\multicolumn{1}{c}{} & \multicolumn{10}{c}{\textbf{CodecFake}} \\
\midrule
Qwen2 audio-chat & 15.92 & 83.42 & 16.24 & 82.91 & 16.74 & 82.33 & 15.38 & 83.91 & 15.04 & 84.23 & 15.86 & 83.36 \\
Qwen2 audio-base & 15.43 & 83.79 & 16.43 & 82.28 & \textbf{17.91} &\textbf{ 81.26} & 16.87 & 81.59 & 16.29 & 81.86 & 16.59 & 82.16 \\
\bottomrule
\end{tabular}
\label{zeroshot}
\end{table*}



\noindent We keep the LLM decoder frozen. A separate decision prompt, (\textit{Determine whether the speech is real or fake. Answer only in one word: ``Real'' or ``Fake''}), is provided to the decoder, which generates the output sequence $Y$. This decision prompt is selected based on initial experiments (Section \ref{experiments} Experimental Results: Zero-shot evaluation of ALMs) across different prompt templates. The decoder output is constrained to either \textit{Real} or \textit{Fake}. The language modeling loss is defined as
\[
\mathcal{L}_{LM}
=
-\sum_{t=1}^{L}
\log p(y_t \mid y_{<t}, g, P_B).
\]
The complete training objective is
\[
\mathcal{L}
=
\lambda_1 \mathcal{L}_{S\text{-}S}
+
\lambda_2 \mathcal{L}_{S\text{-}T}
+
\lambda_3 \mathcal{L}_{LM},
\]
where $\lambda_1, \lambda_2,$ and $\lambda_3$ control the contributions of speech fusion, prompt conditioning, and language generation, respectively. Also, the alignment losses are transformed back to Euclidean space and then added together with the training objective for joint optimization. As the BD losses are loss functions and don't generally add much overhead on parameters. So, the total trainable parameters is approximately 3.75M.



\section{Experiments}
\label{experiments}
\subsection{Training Details and Hyperparameters}
\textbf{\texttt{SATYAM}} is trained in a supervised manner on the ICF training split with only lightweight CNN layers, projection layers, and hyperbolic alignment modules are optimized. Training is performed using AdamW with a learning rate of $1\times10^{-4}$, batch size of 32, for 5 epochs. We keep the values of control parameters of the training objective $\lambda_1$, $\lambda_2$, $\lambda_3$ to be 1, 0.5, and 1 respectively after initial experimentation on the validation set of ICF. More details are given in Appendix \ref{sec:appendix} Additional Training Details.

\subsection{Experimental Results}

We use Accuracy (ACC) and EER as the evaluation metrics for our experiments as used by previous research on speech deepfake detection incudling CF detection \cite{phukan2024heterogeneity, wu24p_interspeech, lu2024codecfake}. All the experiments with ICF has been carried out on CFs generated using test-known set of Indic-SUPERB. Only the experiments for \textbf{Unseen-codec evaluation (clean and noisy test-unknown)} below, we perform on CFs generated using test-unknown (Second evaluation setting as mentioned in Section \ref{sec:indic-codecfake} Generation Pipeline of ICF).

\noindent \textbf{Training on previous benchmark CF dataset and testing on ICF}: We train AASIST on CodecFake \cite{wu24p_interspeech} and then evaluate it zero-shot on ICF. While AASIST performs strongly in-domain on CodecFake (94.21\% ACC / 10.13\% EER, Table~2), its performance drops sharply on ICF to 48.0\% ACC and 40.32\% EER. This large degradation indicates a substantial distribution shift and poor generalization from English-centric CodecFake conditions to Indic scenarios.

\noindent \textbf{Zero-shot Evaluation of ALMs on CF detection}: Gu et al. \cite{10.1145/3746027.3755851} carried out the first study of evaluating ALMs for speech deepfake detection excluding CF detection. They showed the Qwen2-audio ALMs generally perform better than other ALMs. So, we also carried out zero-shot evaluation of Qwen2-audio family on ICF and Codecfake \cite{wu2024codecfake} datasets. The results are presented in Table \ref{zeroshot}. Our results shows that Qwen2-audio-base generally performns better in both datasets as well under different prompt templates and the results with Prompt3 (Used for \textbf{\texttt{SATYAM}} Section \ref{satyam}) being the best. The prompt templates are given in Appendix \ref{sec:appendix} Table \ref{prompts}. Our results reciporcate with the results obtained by Gu et al. \cite{10.1145/3746027.3755851} with Qwen2-audio-base being the best in zero-shot manner. We also carried out further zero-shot analysis of ALMs with different SOTA ALMs such Pengi \cite{deshmukh2023pengi}, Audio Flamingo 2 \cite{ghoshaudio}, Audio Flamingo 3 \cite{ghosh2026audio}, Qwen-audio-chat, and lastly Qwen-audio-base \cite{chu2023qwen}. We carried out these experiments with Prompt 3. The results are presented in Table \ref{tab:indic_codecfake} Zero-shot evaluation of ALMs. We observe that Pengi performed the worst among all the ALMs and this can be due its lack in speech-specific pre-training done. While Qwen2-audio-base being the topmost, however, the results are the overall poor with very less accuracy and high EER. This calls for the need of CF specific training. Also, we believe we are the first study, to the best of our knowledge, to carry out evaluation of ALMs for CF detection.

\begin{table}[hbt!]
\centering
\scriptsize
\setlength{\tabcolsep}{3pt}
\renewcommand{\arraystretch}{1.15}
\begin{tabular}{l|cc|cc}
\toprule
\multirow{2}{*}{\textbf{Method}} &
\multicolumn{2}{c|}{\textbf{ICF}} &
\multicolumn{2}{c}{\textbf{CF}} \\
& {\textbf{ACC \(\uparrow\)}} & {\textbf{EER \(\downarrow\)}} & {\textbf{ACC \(\uparrow\)}} & {\textbf{EER \(\downarrow\)}} \\
\midrule

\multicolumn{5}{c}{\textbf{Zero-shot evalution of ALMs}} \\
\midrule
Pengi & 3.19 & 98.26 & 5.68 & 94.97 \\
Audioflamingo 2 & 5.42 & 97.68 & 8.41 & 92.10 \\
Audioflamingo 3 & 6.98 & 97.21 & 10.22 & 90.85 \\
Qwen-audio-chat & 10.63 & 89.71 & 13.00 & 86.61 \\
Qwen-audio-base &11.17 & 89.23 & 15.82 & 85.74 \\
Qwen2 audio-chat & 12.05 & 88.95 & 16.74 & 82.33 \\
Qwen2 audio-base & 13.41 & 88.57 & 17.91 & 81.26 \\
\midrule
\multicolumn{5}{c}{\textbf{End-to-end method}} \\
\midrule
AASIST & 90.60 & 12.47 & 94.21 & 10.13 \\
\midrule

\multicolumn{5}{c}{\textbf{Pre-Trained Backbone}} \\
\midrule
W-LCCN & 91.98 & 11.89 & 93.38 & 7.92 \\
Wav2vec2-AASIST & 92.50 & 9.62 & 94.45 & 7.29 \\
W + T (LF) & 	92.60 & 9.53 & 94.83 & 7.19 \\
W + T (CA) & 92.65 & 9.48 & 94.99 & 7.01 \\
MiO & 92.80 & 9.04 & 95.11 & 6.49 \\
\midrule

\multicolumn{5}{c}{\textbf{Fine-Tuning ALM}} \\
\midrule
Qwen2 audio-base & 93.19 & 8.34 & 95.55 & 5.60 \\
\midrule
\multicolumn{5}{c}{\textbf{Our Approaches}} \\
\midrule
W + Qwen2-7B & 92.98 & 8.61 & 94.64 & 6.02 \\
T + Qwen2-7B & 93.21 & 8.09 & 95.10 & 5.83 \\
W + T + Qwen2-7B (C) & 93.28 & 7.94 & 95.75 & 4.39 \\
W + T + Qwen2-7B(MA) & 94.01 & 7.02 & 95.31 & 4.07 \\
W + T + Qwen2-7B (E-BD) & 94.93 & 5.39 & 96.47 & 3.68 \\
W + T + Qwen2-7B (H-BD-ST) & 95.78 & 5.14 & 97.22 & 2.69 \\
W + T + Qwen2-7B (H-BD-SS) & 96.11 & 5.02 & 97.34 & 2.42 \\
\textbf{\texttt{SATYAM}} &
\bfseries \cellcolor{blue!25}98.32 & \bfseries \cellcolor{blue!25}3.27 &\cellcolor{blue!25}\textbf{99.11} & \cellcolor{blue!25}\textbf{1.94} \\
\textbf{\texttt{SATYAM}} with Qwen2-1.8B  & \cellcolor{yellow!25}97.14 & \cellcolor{yellow!25}4.53 & \cellcolor{yellow!25}98.32 & \cellcolor{yellow!25}2.11 \\
\bottomrule
\end{tabular}
\caption{Evaluation Results on ICF and Codecfake (CF) \cite{wu24p_interspeech}; ACC stands for accuracy. Blue cells indicate the best (ACC/EER), while yellow cells indicate the second highest results. Scores are in \%; W, T stands for Whisper and TRILLsson respectively; CA, LF stands for Cross-attention and Late-Fusion respectively}
\label{tab:indic_codecfake}
\vspace{-0.5cm}
\end{table}

\noindent \textbf{In-domain Training and Evaluation}: Table~\ref{tab:indic_codecfake} reports the results obtained by training and evaluating on the ICF dataset. We consider AASIST~\cite{wu24p_interspeech}, Wav2vec2-AASIST~\cite{lu2024codecfake}, Whisper-LCNN~\cite{kawa23b_interspeech}, and MiO~\cite{hetero2024} as traditional classifier-based baselines. The corresponding training details are provided in the Appendix \ref{sec:appendix} Additional Training Details. We also add supervised fusion baselines with the audio encoders considered in our study for \textbf{\texttt{SATYAM}}, cross-attention between Whisper and TRILLsson representations (W+T (CA)) and late fusion (W+T (LF)). Both approaches use the same CNN projection module as in \textbf{\texttt{SATYAM}}, whose downstream effectiveness has been shown in prior work \cite{phukan2024heterogeneity}. In late fusion, encoder representations are first processed through the CNN projection module to produce class-level logits, followed by a projection layer and final classification layer as in Sharma et al. \cite{sharma2023late}. For all supervised baselines, the audio encoders were kept frozen and the models were trained using the same configuration as MiO \cite{phukan2024heterogeneity}, a representative encoder fusion framework for speech deepfake detection. Among these approaches, MiO achieves the best performance, highlighting the effectiveness of multi-encoder fusion for CF detection. This observation is consistent with prior findings, where MiO demonstrated SOTA performance for vocoder-based synthetic speech detection using multi-encoder fusion. Further, we add a fine-tuning of ALM baseline (Fine-Tuning ALM). We fine-tune the projection layer of Qwen2-Audio-base (Qwen2-Audio-base uses Whisper audio encoder + Qwen2-7B LLM decoder), selected because it achieved the strongest zero-shot performance in our experiments and has demonstrated effectiveness in prior work \cite{10.1145/3746027.3755851}. For a fair comparison with \textbf{\texttt{SATYAM}}, we followed the same training protocol: both the audio encoders and the LLM decoder were kept frozen, and only the projection layers were trained using identical training settings. \par

We next evaluate our proposed framework, \textbf{\texttt{SATYAM}}, which achieves the best overall performance on the ICF dataset, demonstrating its effectiveness. Fine-tuning Qwen2-Audio-base improves performance over the W+Qwen2-7B setup (which shares the same architecture), likely due to prior exposure to large-scale audio during pretraining; however, \textbf{\texttt{SATYAM}} still achieves the best performance. To better understand the contribution of each component, we further conduct a series of ablation studies with \textbf{\texttt{SATYAM}}, as reported under \emph{Our Approaches} in Table~\ref{tab:indic_codecfake}. All ALM-based variants employ the same conditioning prompt and follow the same training protocol as the full \textbf{\texttt{SATYAM}} model. We first experiment with single-encoder configurations (\textit{W + Qwen2-7B} and \textit{T + Qwen2-7B}). Among these, TRILLsson-based model perform better, reflecting the predominantly paralinguistic nature of speech deepfake detection. We then explore simple concatenation in Euclidean space for both speech--speech and speech--prompt fusion (\textit{W + T + Qwen2-7B (C)}). This is followed by geometry-aware fusion using mobius addition after mapping representations to hyperbolic space in both fusion stages (\textit{W + T + Qwen2-7B (MA)}). Next, we evaluate BD–based alignment in euclidean space for both fusion stages (\textit{W + T + Qwen2-7B (E-BD)}), following prior work~\cite{phukan25d_interspeech}. Finally, we disentangle the role of hyperbolic alignment by applying BD only to speech--speech fusion (\textit{H-BD-SS}) and, in contrast, only to speech--prompt fusion (\textit{H-BD-ST}). These ablations collectively highlight the complementary benefits of hyperbolic geometry and dual-stage alignment in \textbf{\texttt{SATYAM}}. Furthermore, we replace the original decoder with a lightweight LLM, Qwen2-1.8B\footnote{\url{https://huggingface.co/Qwen/Qwen-1_8B}}, and observe a slight decrease in performance compared to the full \textbf{\texttt{SATYAM}} configuration using Qwen2-7B. Nevertheless, this lightweight variant still substantially outperforms the single-encoder configurations with Qwen2-7B, indicating that the quality of the audio encoders constitutes the primary performance bottleneck, consistent with the findings of Li et al.~\cite{li2025dfallm}. Importantly, these results demonstrate that competitive performance can be achieved even without fine-tuning the LLM decoder, further validating the effectiveness and efficiency of the proposed framework. We also present the results of \textbf{\texttt{SATYAM}} across individual Indic-languages in Appendix \ref{sec:appendix} Table \ref{languageeer} showing consistent performance across all the languages. Furthermore, we carried out statistical significance test on our results in Appendix \ref{sec:appendix} Statistical Significance and thus supporting the statistical validation of our obtained results. Similar performance is observed acorss Codecfake \cite{wu24p_interspeech} too, with \textbf{\texttt{SATYAM}} performing the best. Further, we conduct a prompt analysis of \textbf{\texttt{SATYAM}}, including its lightweight variant \textbf{\texttt{SATYAM}} with Qwen2-1.8B, which achieves the second-best overall performance. The results are presented in Appendix \ref{sec:appendix} Prompt Analysis of \textbf{\texttt{SATYAM}}. We observe with varied prompts also through usage of conditioning prompt with \textbf{\texttt{SATYAM}}, we are getting better performance than without usage of conditioning prompt.

\noindent \textbf{Comparison of SOTA models and \texttt{\textbf{SATYAM}} on previous CF benchmark}: We evaluate performance on prior CF detection benchmark, CodecFake by Wu et al. \cite{wu24p_interspeech}. We perform two types of evaluation: (i) in-domain training and testing on CodecFake, and (ii) cross-evaluation between ICF and CodecFake. As a baseline, we consider AASIST as it shown its effectiveness in CF detection \cite{wu24p_interspeech, lu24f_interspeech}. In the in-domain setup, \textbf{\texttt{SATYAM}} outperforms the standard AASIST baseline by a large margin (Table \ref{tab:indic_codecfake}). For cross-benchmark transfer, \texttt{\textbf{SATYAM}} remains robust in both directions, achieving low EER when trained on ICF and tested on CodecFake (3.79\% EER), and when trained on CodecFake and tested on ICF (7.43\% EER). In contrast, AASIST exhibits substantial degradation under the same distribution shifts, reaching 29.81\% EER for ICF$\rightarrow$CodecFake transfer and 40.32\% EER for CodecFake$\rightarrow$ICF transfer.

\noindent \textbf{Language Family}: We further analyze generalization across languages by evaluating \texttt{\textbf{SATYAM}} under (i) random cross-lingual splits and (ii) language-family transfer between the two dominant families in our dataset, namely Dravidian and Indo-European (Appendix \ref{sec:appendix} IndicSUPERB shows the Dravidian and Indo-European languages). Under the random cross-lingual setting, where the model is trained on six randomly selected languages and evaluated on the held-out remaining language, \texttt{\textbf{SATYAM}} maintains low error in both directions (6.34\% and 7.09\% EER). In contrast, AASIST showed much higher EER (26.74\% and 31.11\%). For language-family transfer, we train on Dravidian languages and evaluate on Indo-European languages, and vice versa. \texttt{\textbf{SATYAM}} remains stable under this structured distribution shift, obtaining 7.78\% EER for Dravidian$\rightarrow$Indo-European transfer and 8.48\% EER for Indo-European$\rightarrow$Dravidian transfer. For AASIST, for Indo-European (Train)-> Dravidian (Test), we got EER of 33.45\% and vice-versa, we got EER of 38.73\%. Overall, these results indicate that \texttt{\textbf{SATYAM}} generalizes effectively across unseen languages and even across language families. 

\noindent \textbf{Unseen-codec evaluation (clean and noisy test-unknown)}: To assess robustness under unseen generation conditions, we evaluate on two held-out \texttt{test-unknown} splits: a clean split and a noisy split. The clean split consists of CFs generated using NACs that are unseen during training, as defined in the second evaluation setting in Section~\ref{sec:indic-codecfake} Generation Pipeline of ICF. The noisy split is constructed by generating CFs from the \texttt{test-unknown (noisy)} portion of IndicSUPERB using the same set of unseen NACs as in the clean split.  Across both unseen splits, \texttt{\textbf{SATYAM}} remains reliable, with only a moderate performance degradation when moving from clean to noisy conditions, achieving an EER of 5.23\% on the clean unseen split and 7.41\% on the noisy unseen split. In contrast, AASIST performs substantially worse under the same conditions, with EERs of 14.38\% and 16.29\% on the clean and noisy unseen splits, respectively. These results highlight the robustness of \texttt{\textbf{SATYAM}} to both codec mismatch and adverse acoustic conditions. 

\noindent \textbf{Inference}: \textbf{\texttt{SATYAM}} introduces only lightweight alignment operations (~3.75M parameters) while keeping both audio encoders and the LLM frozen, so inference is dominated by a single backbone forward pass and the hyperbolic mappings add negligible overhead. In practice, W + Qwen2-7B takes 8.00 s on a single-core A100, \textbf{\texttt{SATYAM}} takes 8.18 s, and \textbf{\texttt{SATYAM}} with the lighter decoder takes 6.53 s (averaged over the ICF test set). Notably, \textbf{\texttt{SATYAM}} with the lighter decoder (Qwen2-1.8B) achieves better performance than W + Qwen2-7B, despite sharing a similar architectural (language model decoder) backbone (Qwen2-Audio), which has shown strong performance for speech deepfake detection \cite{10.1145/3746027.3755851}.

\section{Conclusion}  

In this work, we introduced ICF, the first large-scale benchmark comprising real and NAC-synthesized speech across multiple Indic languages, diverse speaker profiles, and multiple NAC types. Our experiments show that SOTA CF detectors trained on English-centric datasets perform poorly on ICF underscoring the effect of changes in linguistic distribution. We further conducted a systematic zero-shot evaluation of SOTA ALMs on ICF, revealing consistent performance degradation despite their effectiveness on other speech tasks. To overcome these limitations, we proposed \textbf{\texttt{SATYAM}}, a hyperbolic ALM tailored for CF detection in Indic languages. By leveraging dual-stage hyperbolic Bhattacharya distance to align semantic and prosodic speech representations and subsequently integrate speech and textual prompts, \textbf{\texttt{SATYAM}} effectively models hierarchical relationships within and across modalities. Extensive evaluations demonstrate that \textbf{\texttt{SATYAM}} consistently outperforms competitive end-to-end and ALLM-based baselines on the ICF benchmark. 

\section*{Limitations}

One limitation of our work is that we consider a single LLM decoder family. However, prior studies have shown that the choice of LLM decoder has a relatively limited impact on performance in audio LLM pipelines~\cite{li2025dfallm}, a trend that is also reflected in our experimental results. In addition, our framework employs two audio encoders; while both are SOTA and well-suited for CF detection, alternative encoder choices may lead to minor performance variations. In future work, we plan to explore a broader range of LLM decoder architectures as well as additional audio encoders to further assess the generality of the proposed framework.

\section*{Ethical considerations}
This work is motivated by the need to improve the robustness of CF detection in low-resource and multilingual settings. We do not collect any new human-subject recordings; the ICF dataset is constructed by applying NACs to the openly available IndicSUPERB corpus, in accordance with its original licenses and usage terms. Although our contributions are defensive in nature, we acknowledge that insights into codec artifacts and detector behavior could potentially be misused. We strongly discourage any unethical or malicious use of the ICF dataset. The benchmark and models introduced in this work are intended strictly for research purposes.




\bibliography{custom}

\section{Appendix}

\label{sec:appendix}

\begin{table}[!hbt]
\centering
\setlength{\tabcolsep}{16pt}
\caption{Language-wise EER (\%)}
\label{tab:kathbath_acc_eer}
\begin{tabular}{clcc}
\toprule
\# & Language  & EER \\
\midrule
1  & Bengali   &  2.95 \\
2  & Gujarati  &  2.98 \\
3  & Kannada   &   2.69\\
4  & Hindi     &   2.34\\
5  & Malayalam &   3.64\\
6  & Marathi   &   2.52\\
7  & Odia      &   2.85\\
8  & Punjabi   &  3.16 \\
9  & Sanskrit  &   3.41\\
10 & Tamil     &   4.11\\
11 & Telugu    &   4.01\\
12 & Urdu      &   3.38\\
\bottomrule
\end{tabular}
\label{languageeer}
\end{table}


\begin{table}[hbt!]
\centering
\begin{tabular}{p{0.95\linewidth}}
\toprule
\begin{itemize}
  \item Prompt 1: \textit{Is this speech real or fake? Reply with one word only: ``Real'' or ``Fake''.}
  \item Prompt 2: \textit{What is the authenticity of this speech? Answer ``Fake'' or ``Real''.}
  \item Prompt 4:\textit{Can you determine if this speech is fake or real? Answer ``Fake'' or ``Real''.}
  \item Prompt 5:\textit{Is this a real speech recording? Answer ``Fake'' or ``Real''.}
  \item Prompt 6:\textit{Is this a AI-generated speech sample? Answer ``Fake'' or ``Real''.}
\end{itemize}
\\
\bottomrule
\end{tabular}
\caption{Decision Prompt Templates}
\label{prompts}
\end{table}

\begin{figure*}[hbt!]
  \centering
  \includegraphics[width=0.89\textwidth]{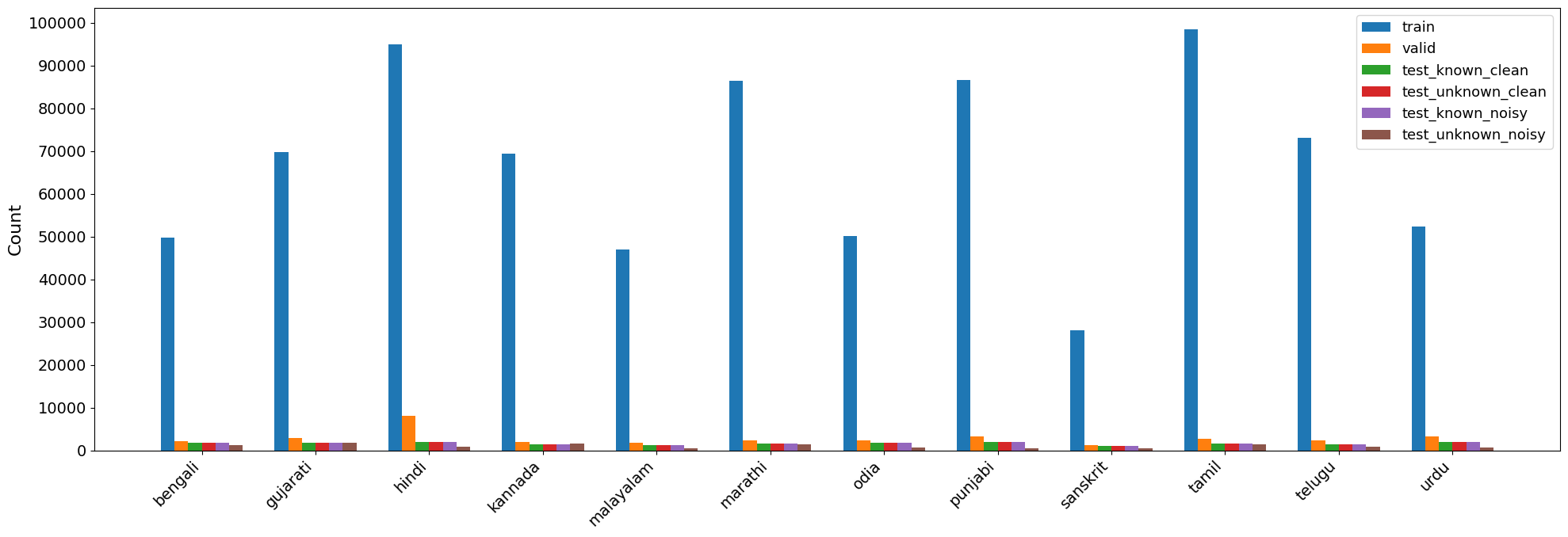}
  \caption{IndicSUPERB data distribution across Train, Val, and Test sets for different Indic languages}
  \label{fig:indicdistri}
\end{figure*}

\subsection{IndicSUPERB}

IndicSUPERB consists of languages: Bengali, Gujarati, Hindi, Kannada, Malayalam, Marathi, Odia, Punjabi, Sanskrit, Tamil, Telugu, and Urdu. Kannada, Malayalam, Tamil, and Telugu falls under Dravidian language family whereas the others languages falls under Indo-European language family. The distribution plot of IndicSUPERB across train, val, and test splits is given in Figure \ref{fig:indicdistri}.

\begin{table*}[hbt!]
\centering
\scriptsize
\setlength{\tabcolsep}{4pt}
\renewcommand{\arraystretch}{1.15}
\begin{tabular}{l|cc|cc|cc|cc|cc|cc}
\toprule
\multicolumn{1}{c}{} & \multicolumn{10}{c}{\textbf{ICF}} \\
\midrule
\textbf{Prompt}
& \multicolumn{2}{c|}{\textbf{Prompt1}}
& \multicolumn{2}{c|}{\textbf{Prompt2}}
& \multicolumn{2}{c|}{\textbf{Prompt3}}
& \multicolumn{2}{c|}{\textbf{Prompt4}}
& \multicolumn{2}{c|}{\textbf{Prompt5}}
& \multicolumn{2}{c}{\textbf{Prompt6}} \\
& \textbf{ACC \(\uparrow\)} & \textbf{EER \(\downarrow\)}
& \textbf{ACC \(\uparrow\)} & \textbf{EER \(\downarrow\)}
& \textbf{ACC \(\uparrow\)} & \textbf{EER \(\downarrow\)}
& \textbf{ACC \(\uparrow\)} & \textbf{EER \(\downarrow\)}
& \textbf{ACC \(\uparrow\)} & \textbf{EER \(\downarrow\)}
& \textbf{ACC \(\uparrow\)} & \textbf{EER \(\downarrow\)} \\
\midrule
\textbf{\texttt{SATYAM}} with Qwen2-1.8B     & 88.13 & 9.63 & 88.29 & 9.36 & 88.87 & 8.18 & 88.85 & 9.04 & 87.52 & 9.49 & 88.33 & 9.14 \\
\textbf{\texttt{SATYAM}} & 92.00 & 7.10 & 90.08 & 7.37 & 91.35 & 7.08 & 92.46 & 6.39 & 89.75 & 7.95 & 91.13 & 7.18 \\
\midrule
\multicolumn{1}{c}{} & \multicolumn{10}{c}{\textbf{CodecFake \cite{wu24p_interspeech}}} \\
\midrule
\textbf{\texttt{SATYAM}} with Qwen2-1.8B    & 90.54 & 6.42 & 90.42 & 5.93 & 92.01 & 5.61 & 90.06 & 5.72 & 89.75 & 6.32 & 90.56 & 6.08 \\
\textbf{\texttt{SATYAM}}  & 92.00 & 5.67 & 92.05 & 5.38 & 93.83 & 5.27 & 92.64 & 5.18 & 92.49 & 5.87 & 92.60 & 5.47 \\
\bottomrule
\end{tabular}
\caption{Evaluation scores with different prompt templates and without the conditioning prompt; ACC stands for Accuracy; All the scores are in \%}
\label{singleprompt}
\end{table*}

\begin{table*}[hbt!]
\centering
\scriptsize
\setlength{\tabcolsep}{4pt}
\renewcommand{\arraystretch}{1.15}
\begin{tabular}{l|cc|cc|cc|cc|cc|cc}
\toprule
\multicolumn{1}{c}{} & \multicolumn{10}{c}{\textbf{ICF}} \\
\midrule
\textbf{Prompt}
& \multicolumn{2}{c|}{\textbf{Prompt1}}
& \multicolumn{2}{c|}{\textbf{Prompt2}}
& \multicolumn{2}{c|}{\textbf{Prompt3}}
& \multicolumn{2}{c|}{\textbf{Prompt4}}
& \multicolumn{2}{c|}{\textbf{Prompt5}}
& \multicolumn{2}{c}{\textbf{Prompt6}} \\
& \textbf{ACC \(\uparrow\)} & \textbf{EER \(\downarrow\)}
& \textbf{ACC \(\uparrow\)} & \textbf{EER \(\downarrow\)}
& \textbf{ACC \(\uparrow\)} & \textbf{EER \(\downarrow\)}
& \textbf{ACC \(\uparrow\)} & \textbf{EER \(\downarrow\)}
& \textbf{ACC \(\uparrow\)} & \textbf{EER \(\downarrow\)}
& \textbf{ACC \(\uparrow\)} & \textbf{EER \(\downarrow\)} \\
\midrule
\textbf{\texttt{SATYAM}} with Qwen2-1.8B      & 94.87 & 5.87 & 94.63 & 6.16 & 97.14 & 4.53 & 95.38 & 5.29 & 94.46 & 5.96 & 95.02 & 5.68 \\
\textbf{\texttt{SATYAM}} & 97.35 & 3.62 & 96.78 & 3.89 & 98.32 & 3.27 & 97.64 & 3.35 & 96.45 & 3.97 & 97.31 & 3.62 \\
\midrule
\multicolumn{1}{c}{} & \multicolumn{10}{c}{\textbf{CodecFake \cite{wu24p_interspeech}}} \\
\midrule
\textbf{\texttt{SATYAM}} with Qwen2-1.8B      & 97.47 & 2.54 & 96.32 & 2.54 & 98.32 & 2.11 & 96.89 & 2.64 & 96.22 & 2.72 & 96.62 & 2.63 \\
\textbf{\texttt{SATYAM}}  & 97.26 & 2.21 & 97.49 & 2.08 & 99.11 & 1.94 & 98.43 & 1.99 & 97.80 & 2.00 & 98.02 & 2.04 \\
\bottomrule
\end{tabular}
\caption{Evaluation scores with different prompt templates and without the conditioning prompt; The conditioning prompt is kept same for all the different versions of the decision prompt; ACC stands for Accuracy; All the scores are in \%}
\label{doubleprompt}
\end{table*}

\subsection{More Information about NACs}

\textbf{DAC}\footnote{\url{https://huggingface.co/descript/dac_16khz}}: It is high-fidelity VQ-GAN codec with RVQ and a mix of adversarial and multi-scale spectral losses to improve reconstruction. \par

\noindent \textbf{Encodec}\footnote{\url{https://huggingface.co/facebook/encodec_24khz}}: It is a real-time, high-fidelity NAC with a convolutional encoder–decoder and RVQ. It combines time- and frequency-domain losses with spectrogram-based adversarial training. \par

\noindent \textbf{SoundStream}\footnote{\url{https://github.com/haydenshively/SoundStream}}: A low-bitrate RVQ codec with multi-scale STFT discriminators, designed to balance fidelity and compression. It supports 3–18 kbps. \par

\noindent \textbf{SpeechTokenizer}\footnote{\url{https://github.com/ZhangXInFD/SpeechTokenizer.git}}: It is a unified tokenizer that bridges semantic and acoustic cues via hierarchical RVQ layers. \par

\noindent \textbf{FunCodec}\footnote{\url{https://github.com/modelscope/FunCodec}}: It is RVQ-based NAC enhanced with semantic augmentation and adversarial strategies. \par

\noindent \textbf{AudioDec}\footnote{\url{https://github.com/facebookresearch/AudioDec}}: It high-fidelity NAC trained in two stages: metric losses for stability, followed by decoder-only adversarial finetuning for realism.  \par

\noindent \textbf{SNAC}\footnote{\url{https://huggingface.co/hubertsiuzdak/snac_44khz}}: It is an RVQ extension with hierarchical quantizers at multiple time scales. \par

\noindent \textbf{MIMI}\footnote{\url{https://huggingface.co/kyutai/mimi}}: It is a encoder-decoder based high fidelity NAC with quantization trained in a end-to-end manner. 

\subsection{More Information about the Audio Encoders}

Whisper \footnote{\tiny\url{https://huggingface.co/openai/whisper-base}} is a transformer-based encoder--decoder architecture trained on 96 languages in a multi-task setting. We use only the encoder to extract representations, obtaining a 512-dimensional embedding after average pooling. Since Whisper is primarily trained for automatic speech recognition, it effectively captures semantic and linguistic information in speech. Following this, we employ TRILLsson\footnote{\tiny\url{https://www.kaggle.com/models/google/trillsson}}, a distilled self-supervised model pretrained for paralinguistic speech processing. TRILLsson has demonstrated strong performance in tasks such as speech emotion recognition, speaker identification, and speech deepfake detection. We extract a 1024-dimensional representation from TRILLsson. Both Whisper and TRILLsson are kept frozen during training, and all input utterances are resampled to 16\,kHz before being passed into the encoders.

\subsection{Additional Training Details}

We use four-core A100 for training our models. Also, the end-to-end baselines and the baselines with pre-trained backbone are trained for 20 epochs with same learning rate and batch size as of \textbf{\texttt{SATYAM}}.

\subsection{Prompt analysis of \texttt{SATYAM}}

We use the same set of decision prompts employed in the zero-shot evaluation of AMs (prompt templates are provided in the Table \ref{prompts}).Table~\ref{singleprompt} reports results obtained using only the decision prompts, while Table~\ref{doubleprompt} reports results when both the conditioning prompt and the decision prompt are used. Across both model variants, we observe that Prompt3, which is adopted in the final \texttt{\textbf{SATYAM}} configuration, consistently yields the best performance. Moreover, incorporating the conditioning prompt leads to systematically improved results across all prompt templates, highlighting the effectiveness of prompt-conditioned alignment in the proposed framework.

\subsection{Statistical Significance}
\label{sec:stat_sig}

We assess statistical significance using a two-sided McNemar’s test on paired predictions from the same test set. McNemar’s test is a preferred statistical signifance used in previous deepfake detection work \cite{batra2025melody}. SATYAM shows statistically significant improvements over AASIST, the strongest SSL baseline (MiO), and SATYAM with Qwen2-1.8B on ICF (\textbf{$p<0.001$} for all comparisons), confirming that the observed gains are statistically significant.

\end{document}